\documentclass[sort&compress,final]{aipproc}
\layoutstyle{6x9}
\begin{document}\title{Systematic errors of bound-state parameters
obtained with SVZ sum rules}
\author{Wolfgang Lucha}{address={HEPHY, Austrian Academy of
Sciences, Nikolsdorfergasse 18, A-1050 Vienna, Austria}}\iftrue
\author{Dmitri Melikhov$^{*,}$}{address={Institute of Nuclear
Physics, Moscow State University, 119992, Moscow, Russia}}
\author{Silvano Simula}{address={INFN, Sezione di Roma 3, Via
della Vasca Navale 84, I-00146, Roma, Italy}}\fi

\begin{abstract}We study systematic errors of the ground-state
parameters obtained by Shifman--Vainshtein--Zakharov (SVZ) sum
rules, making use of the harmonic-oscillator potential model as an
example. In this case, one knows the exact solution for the
polarization operator, which allows one to obtain both the OPE to
any order and the parameters (masses and decay constants) of the
bound states. We determine the parameters of the ground state
making use of the standard procedures of the method of sum rules,
and compare the obtained results with the known exact values. We
show that in the situation when the continuum contribution to the
polarization operator is not known and is modelled by an effective
continuum, the method of sum rules does not allow to control the
systematic uncertainties of the extracted ground-state parameters.
\end{abstract}\maketitle

A QCD sum-rule calculation of hadron parameters \cite{svz}
involves two steps: (i) one calculates the operator product
expansion (OPE) series for a relevant correlator, and (ii) one
extracts the parameters of the ground state by a numerical
procedure. Each of these steps leads to uncertainties in the final
result.

The first step lies fully within QCD and allows a rigorous
treatment of the uncertainties: the correlator in QCD is not known
precisely (because of uncertainties in quark masses, condensates,
$\alpha_s$, radiative corrections, etc), but the corresponding
errors in the correlator may be systematically controlled (at
least in principle).

The second step lies beyond QCD and is more cumbersome: even if
several terms of the OPE for the correlator were known precisely,
the hadronic parameters might be extracted by a sum rule only
within some error, which may be treated as a systematic error of
the method. It is useful to recall that a successful extraction of
the hadronic parameters by a sum rule is not guaranteed: as
noticed already in the classical papers \cite{svz,nsvz}, the
method may work in some cases and fail in others; moreover, error
estimates (in the mathematical sense) for the numbers obtained by
sum rules may not be easily provided --- e.g., according to
\cite{svz}, any value obtained by varying the parameters in the
sum-rule stability region has equal probability. However, for many
applications of sum rules, especially in flavor physics, one needs
rigorous error estimates of the theoretical results for comparing
theoretical predictions with the experimental data. Systematic
errors of the sum-rule results are usually estimated by varying
the Borel parameter and the continuum threshold within some ranges
and are believed to be under control.

Here we present the results of our recent analysis of the
systematic uncertainties of the sum-rule procedure \cite{sr_lms}.
To this end, a quantum-mechanical harmonic-oscillator (HO)
potential model is a perfect tool: in this model both the spectrum
of bound states (masses and wave functions) and the exact
correlator (and hence its OPE to any order) are known precisely.
Therefore, one may apply the sum-rule machinery for extracting
parameters of the ground state and test the accuracy of the
extracted values by comparing with the exact known results. In
this way the accuracy of the method can be probed.

\begin{figure}[t]
\includegraphics[width=10cm]{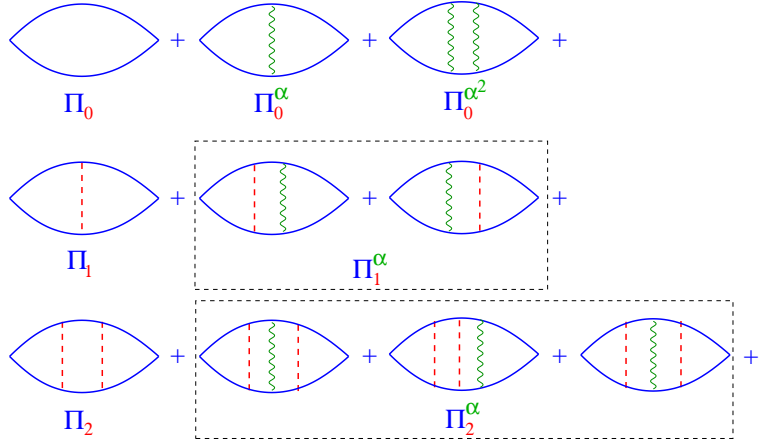}
\caption{\label{Fig:01}The expansion of the polarization operator
$\Pi(p^2)$ in powers of the interaction: wavy lines (green)
represent the Coulomb-part exchange, dashed lines (red) represent
the confining-part exchange.}
\end{figure}

\vspace{.2cm} \noindent{\bf The model.} To illustrate the
essential features of the QCD calculation, let us consider a
non-relativistic model with a potential containing both a Coulomb
and a confining part:
\begin{eqnarray}
H=H_0+V(r), \quad H_0=\vec p^2/2m, \quad V(r)={\alpha_s}/{r}+V_{\rm conf}.
\end{eqnarray}
The polarization operator $\Pi(E)$ is defined by the full Green function
$G(E)=(H-E)^{-1}$ as follows:
\begin{eqnarray}
\label{piE}
\Pi(E)=\left(2\pi/m\right)^{3/2}
\langle \vec r_f=0|G(E)|\vec r_i=0\rangle.
\end{eqnarray}
The full Green function satisfies the Lippmann--Schwinger operator
equation
\begin{eqnarray}
G^{-1}(E)-G_0^{-1}(E)=V,
\end{eqnarray}
which may be solved by constructing the expansion in powers of the
interaction $V$:
\begin{eqnarray}
\label{ls} G(E)=G_0(E)-G_0(E)VG_0(E)+\cdots.
\end{eqnarray}
Respectively, for the polarization operator $\Pi(E)$ one may
construct the expansion shown in Fig.~\ref{Fig:01}. If the
confining potential is known, one can calculate each term of the
expansion. If one does not know the confining potential precisely,
an explicit calculation is not possible. In this case one can
explicitly calculate the contribution
$\Pi_0+\Pi_0^{\alpha}+\Pi_0^{\alpha^2}+\cdots$ of those diagrams
which do not contain the confining potential (first line in
Fig.~\ref{Fig:01}) and parametrize other diagrams, in which the
confining potential appears, taking into account the symmetries of
the theory. In QCD such diagrams are parametrized in terms of
condensates and radiative corrections to them.

In what follows we make a further simplification of the model and
switch off the Coulomb potential, which we, in principle, know how
to deal with. We shall retain only the confining potential, choose
it as the HO potential
\begin{eqnarray}
\label{1.1}
V(r)={m\omega^2 \vec r^2}/{2}, \qquad r=|\vec r|,
\end{eqnarray}
and study $\Pi(\mu)$ [the Borel transform of the polarization
operator $\Pi(E)$] which gives the evolution operator in the
imaginary time $1/\mu$:
\begin{eqnarray}
\Pi(\mu)=\left(2\pi/m\right)^{3/2}
\langle \vec r_f=0|\exp(- H/\mu)|\vec r_i=0\rangle.
\end{eqnarray}
For the HO potential (\ref{1.1}), the exact analytic expression
for $\Pi(\mu)$ is known \cite{nsvz}:
\begin{eqnarray}
\label{piexact}
\Pi(\mu)=\left(\frac{\omega}{\sinh(\omega/\mu)}\right)^{3/2}.
\end{eqnarray}
Expanding this expression in inverse powers of $\mu$, we get the
OPE series for $\Pi(\mu)$:
\begin{eqnarray}
\label{piope}
\Pi_{\rm OPE}(\mu)\equiv \Pi_{0}(\mu)+\Pi_{1}(\mu)+\Pi_{2}(\mu)+\cdots=
\mu^{3/2}
\left[1-\frac{\omega^2}{4\mu^2}+\frac{19}{480}\frac{\omega^4}{\mu^4}
+\cdots \right],
\end{eqnarray}
and higher coefficients may be obtained from (\ref{piexact}).
Each term of this expansion may be also calculated from (\ref{piE}) and (\ref{ls}),
with $\Pi_0$ corresponding to $G_0$.

The ``phenomenological'' representation for $\Pi(\mu)$ is obtained
by using the basis of hadron eigenstates of the model, namely
\begin{eqnarray}
\label{piphen1}
\Pi(\mu)=\sum_{n=0}^\infty R_n \exp(-E_n/\mu),
\end{eqnarray}
with $E_n$ the energy of the $n$-th bound state and $R_n$
(the square of the leptonic decay constant of the $n$-th bound state)
given by
\begin{eqnarray}
R_n=(2\pi/m)^{3/2}|\Psi_n(\vec r=0)|^2.
\end{eqnarray}
For the lowest states, one finds from (\ref{piexact})
\begin{eqnarray}
\label{E0} E_0=\frac{3}{2}\omega,\; R_0=2\sqrt{2}\omega^{3/2},
\quad E_1=\frac{7}{2}\omega,\;
R_1=3\sqrt{2}\omega^{3/2},\quad\ldots.
\end{eqnarray}

\begin{figure}[t]\begin{tabular}{cc}
\includegraphics[width=7.4cm]{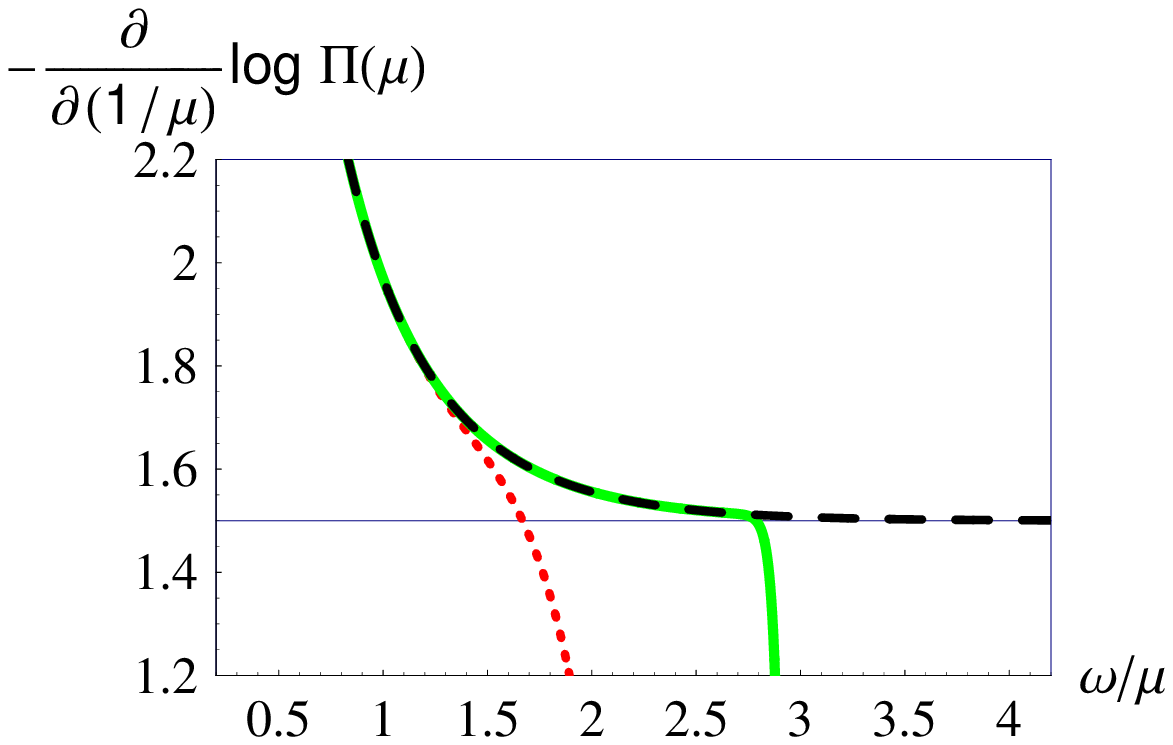}&
\includegraphics[width=7.4cm]{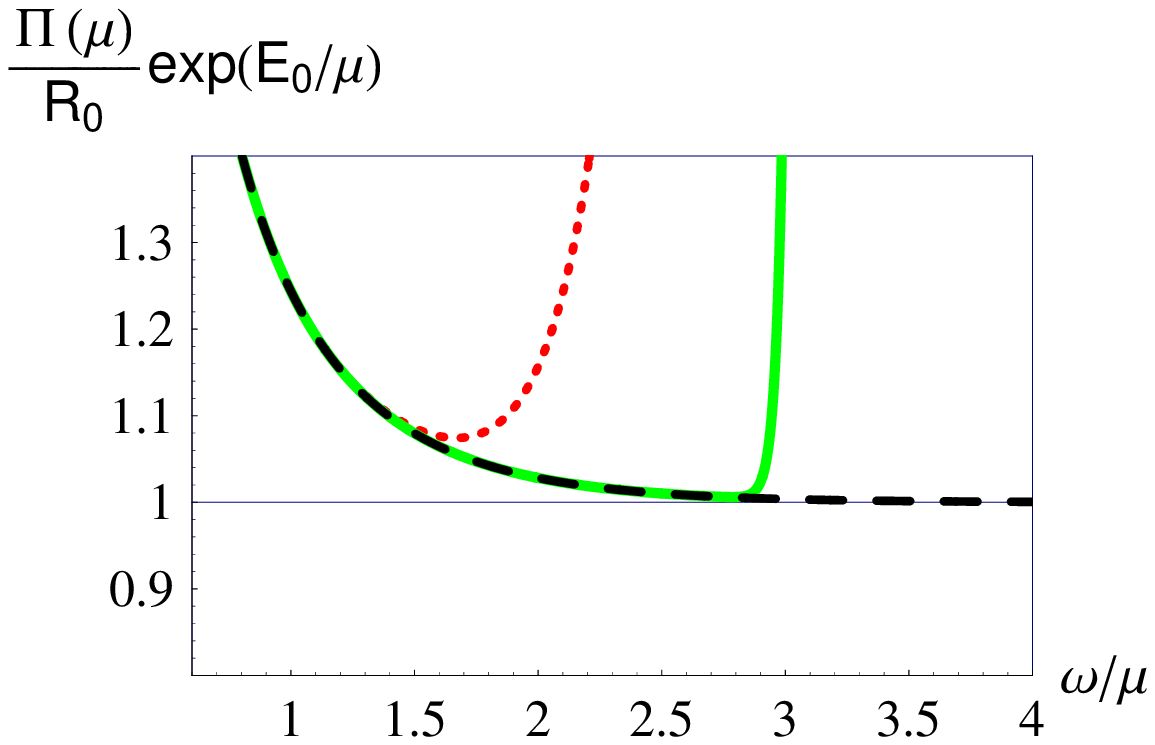}\end{tabular}
\caption{\label{Fig:02} Dashed (black) line -- the calculation for
the exact $\Pi(\mu)$; dotted (red) line -- the calculation making
use of the OPE for $\Pi(\mu)$ with 4 power corrections, solid
(green) line -- the same OPE with 100 power corrections.}
\end{figure}

Now, imagine we know $\Pi(\mu)$ numerically and want to extract
the parameters of the ground state $E_0$ and $R_0$. At large
values of the Euclidean time $1/\mu$ the polarization operator is
dominated by the ground state. Therefore, one has a plateau in
$-\frac{\partial}{\partial \mu}\log \Pi(\mu)$ and in the product
$\exp(E_0/\mu)\Pi(\mu)$ (Fig.~\ref{Fig:02}). In practice, working
at sufficiently large values of $1/\mu$, one constructs
$-\frac{\partial}{\partial \mu}\log \Pi(\mu)$ and, after verifying
that the plateau has already been reached, determines $E_0$. Then
one studies $\Pi(\mu)\exp(E_0/\mu)$ and reads~off~$R_0$.
In principle, it is possible to reach the plateau if one retains
sufficiently many terms of the truncated OPE: the radius of
convergence of the truncated OPE to the exact function $\Pi(\mu)$
increases with the increase of the number of terms taken into
account, and, e.g., for $N>100$ the range of convergence already
reaches the plateau (Fig.~\ref{Fig:02}).

In practice, however, one can calculate only the first few terms
and, consequently, the radius of convergence is rather small. For
instance, with 4 terms the convergence region extends only up to
values of $1/\mu$ for which the ground state gives 85\% of the
correlator.

The method of SVZ sum rules aims at extracting the parameters of
the ground state from the first few terms of the OPE, e.g. in the
region where the contribution of higher states is by far not
negligible.

\vspace{0.3cm}\noindent{\bf Sum rule.} The sum rule claims the
equality of the correlator calculated in the ``quark'' basis and
in the hadron basis:
\begin{eqnarray}
&&R_0 \exp(-{E_0}/{\mu})+\int\limits_{z_{\rm cont}}^\infty dz
\rho_{\rm phen}(z)\exp(-{z}/{\mu})\nonumber\\
&&=\int\limits_{0}^{\infty}dz \rho_0(z)\exp(-{z}/{\mu}) +\mu^{3/2}
\left[ -\frac{\omega^2}{4\mu^2}
+\frac{19}{480}\frac{\omega^4}{\mu^4} +\cdots\right].\label{sr}
\end{eqnarray}
Following \cite{nsvz}, we use explicit expressions for the power
corrections, but for the zero-order free-particle term we use its
expression in terms of the spectral integral. The reason for this
will become clear in a few lines.

Let us introduce the effective continuum threshold $z_{\rm
eff}(\mu)$, different from the physical $\mu$-independent
continuum threshold $z_{\rm cont}$, by the relation
\begin{eqnarray}
\label{zeff}
\Pi_{\rm cont}(\mu)=
\int\limits_{z_{\rm cont}}^\infty dz \,\rho_{\rm phen}(z)\,\exp(-z/\mu)=
\int\limits_{z_{\rm eff}(\mu)}^\infty dz\, \rho_{0}(z)\,\exp(-z/\mu).
\end{eqnarray}
Generally speaking, the spectral densities $\rho_{\rm phen}(z)$
and $\rho_{0}(z)$ are different functions, so the two sides of
(\ref{zeff}) may be equal to each other only if the effective
continuum threshold depends on $\mu$. In our model, we can
calculate $\Pi_{\rm cont}$ precisely, as the difference between
the known exact correlator and the known ground-state
contribution, and therefore we can obtain the function $z_{\rm
eff}(\mu)$ by solving (\ref{zeff}). In the general case of a
sum-rule analysis, the effective continuum threshold is not known
and is one of the essential fitting parameters.

Making use of (\ref{zeff}), we rewrite now the sum rule (\ref{sr})
in the form
\begin{eqnarray}
\label{sr2}
R_0 \exp({-{E_0}/\mu})=\Pi(\mu,z_{\rm eff}(\mu)),
\end{eqnarray}
where the cut correlator $\Pi(\mu,z_{\rm eff}(\mu))$ reads
\begin{eqnarray}
\label{cut}
\Pi(\mu,z_{\rm eff}(\mu))\equiv
\frac{2}{\sqrt{\pi}}\int\limits_{0}^{z_{\rm eff}(\mu) }dz \sqrt{z}\exp(-z/\mu)
+\mu^{3/2}
\left[
-\frac{\omega^2}{4\mu^2}
+\frac{19}{480}\frac{\omega^4}{\mu^4}+\cdots\right].
\end{eqnarray}
As is obvious from (\ref{sr2}), the cut correlator satisfies the equation
\begin{eqnarray}
\label{e0a}
-\frac{d}{d(1/\mu)}\log \Pi(\mu,z_{\rm eff}(\mu)) =E_0.
\end{eqnarray}
The cut correlator is the actual quantity which governs the
extraction of the ground-state parameters.

The sum rule (\ref{sr2}) allows us to restrict the structure of
the effective continuum threshold $z_{\rm eff}(\mu)$. Let us
expand both sides of (\ref{sr2}) near $\omega/\mu=0$. The l.h.s.\
contains only even powers of $\sqrt{\omega/\mu}$; power
corrections on the r.h.s. contain only odd powers of
$\sqrt{\omega/\mu}$. In order that both sides match each other,
the effective continuum threshold cannot be constant but should be
a power series of the parameter $\sqrt{\omega/\mu}$:
\begin{eqnarray}
\label{zeff2}
z_{\rm eff}(\mu)=\omega
\left[\bar{z}_0+\bar{z}_1\sqrt{\frac{\omega}{\mu}}
+\bar{z}_2\frac{\omega}{\mu}+\cdots\right].
\end{eqnarray}
Inserting this series in (\ref{sr2}) and expanding the integral on
the r.h.s., we obtain an infinite chain of equations emerging at
different orders of $\sqrt{\omega/\mu}$. These equations allow us
to obtain for any $E_0$ and $R_0$ within a broad range of values a
solution $z_{\rm eff}(\mu,R_0,E_0)$ which {\it exactly} solves the
sum rule (\ref{sr}). Therefore, in a limited range of $\mu$ the
OPE alone cannot say much about the ground-state parameters. What
really matters is the continuum contribution, or, equivalently,
$z_{\rm eff}(\mu)$. Without constraints on the effective continuum
threshold the results obtained from the OPE are not restrictive.

The approximate extraction of $E_0$ and $R_0$ worked out in a
limited range of values of $\mu$ becomes possible only by
constraining $z_{\rm eff}(\mu)$. If the constraints are realistic
and turn out to reproduce with a reasonable accuracy the exact
$z_{\rm eff}(\mu)$, then the approximate procedure works well. If
a good approximation is not found, the approximate procedure fails
to reproduce the true value. Anyway, the accuracy of the extracted
value is difficult to be kept under control. This conclusion is
quite different from the results of QCD sum rules presented in the
literature (see e.g.~the review \cite{ck}). We shall demonstrate
that a typical sum-rule analysis contains additional explicit or
implicit assumptions and criteria for extracting the parameters of
the ground state. Whereas these assumptions may lead to good
central values of hadron parameters, the accuracy of the extracted
values cannot be controlled within the method of QCD sum rules.

\begin{figure}[t]\begin{tabular}{cc}
\includegraphics[width=6.5cm]{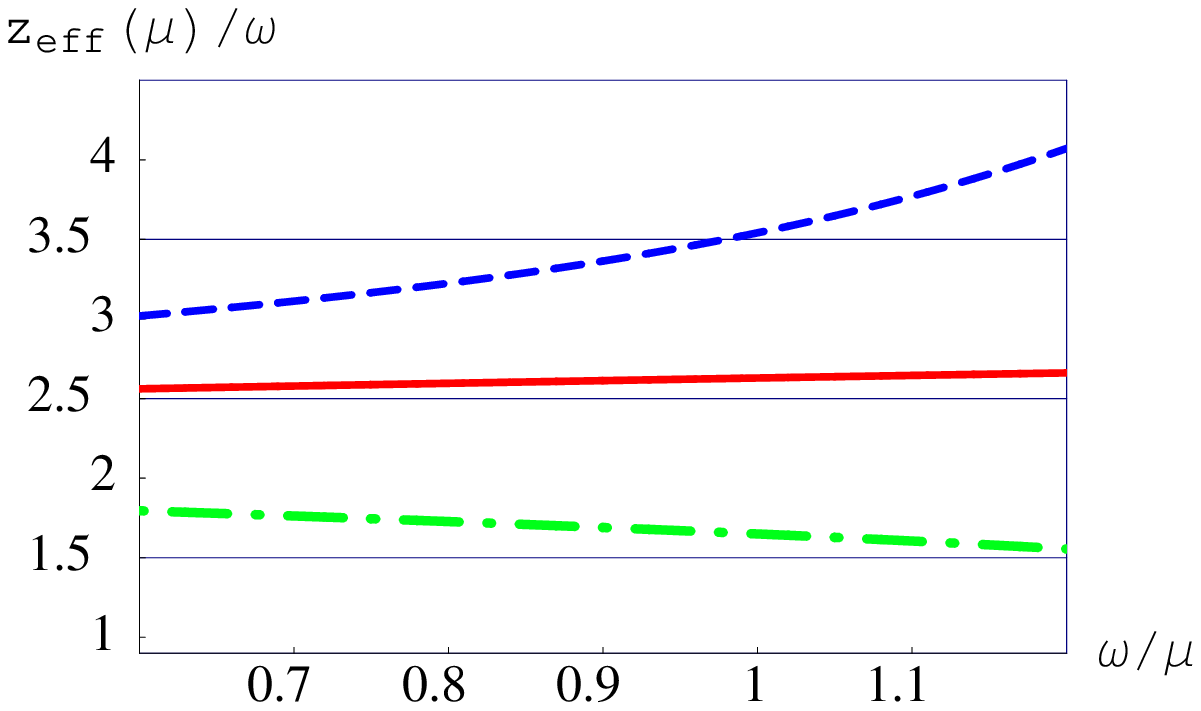}&\hspace{.2cm}
\includegraphics[width=6.5cm]{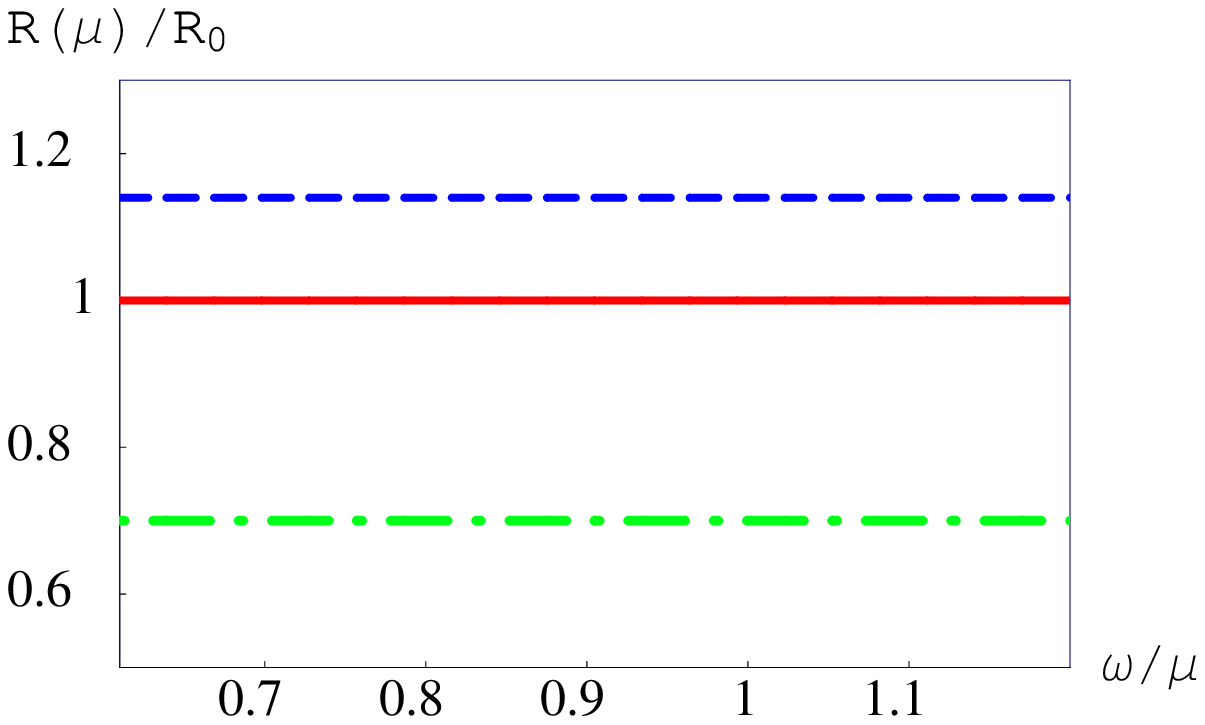}\\\end{tabular}
\caption{\label{Fig:2}Different choices of the effective continuum
threshold $z_{\rm eff}(\mu)$ (a) and the corresponding $R(\mu)$
(b): 1 [solid (red) line] the exact effective continuum threshold
as obtained by a numerical solution of (\ref{fit}) for $E=E_0$ and
$R=R_0$. 2 [long-dashed (blue) line] the effective continuum
threshold obtained by solving the sum rule (\ref{fit}) for $E=E_0$
and $R=0.7 R_0$, 3 [dash-dotted (green) line] same as line 2, but
for for $E=E_0$ and $R=1.15 R_0$.}\end{figure}

\vspace{0.3cm}\noindent{\bf Numerical analysis.} In practice, one
knows only the first few terms of the OPE, so one must stay in a
region of $\mu$ bounded from below to guarantee that the truncated
OPE series reproduces the exact correlator within a controlled
accuracy. The ``fiducial'' \cite{svz} range of $\mu$ is the range
where, on the one hand, the OPE reproduces the exact expression
better than some given accuracy (e.g., within 0.5\%) and, on the
other hand, the ground state is expected to give a sizable
contribution to the correlator. If we include the first three
power corrections, $\Pi_1$, $\Pi_2$, and $\Pi_3$, then the
fiducial region lies at $\omega/\mu<1.2$ (see Fig.~\ref{Fig:2}).
Since we know the ground-state parameters, we fix
$\omega/\mu>0.7$, where the ground state gives more than 60\% of
the full correlator. So our working range is $0.7<\omega/\mu<1.2$.

Obviously, if one knows the continuum contribution with a
reasonable accuracy, one can extract the resonance parameters from
the sum rule (\ref{sr}). We shall be interested, however, in the
situation when the hadron continuum is not known, which is a
typical situation in heavy-hadron physics and in studying
properties of exotic hadrons. Can we still extract the
ground-state parameters?

We shall seek the (approximate) solution to the equation
\begin{eqnarray}
\label{fit}
R \exp({-{E}/\mu})+\int\limits_{z_{\rm
eff}(\mu)}^\infty dz \rho_0(z) \exp(-z/\mu) =\Pi_{\rm OPE}(\mu)
\end{eqnarray}
in the range $0.7<\omega/\mu<1.2$. Hereafter, we denote by $E$ and
$R$ the values of the ground-state parameters as extracted from
the sum rule (\ref{fit}). The notations $E_0$ and $R_0$ are
reserved for the known exact values.

\vspace{.2cm}\noindent\underline{\it $\mu$-dependent effective
continuum threshold.} In many interesting cases the ground-state
energy can be determined, for instance, from experiment. However,
fixing the ground-state energy $E$ equal to its known value $E_0$
does not help: for any $R$ within a broad range one can still find
a solution $z_{\rm eff}(\mu, R)$ which solves the sum rule
(\ref{fit}) exactly (Fig.~\ref{Fig:2}). In order to extract $R$,
one needs to constrain the effective continuum threshold;
moreover, this constraint determines the actual value of $R$ which
one obtains from the sum rule. If one requires, e.g., $z_{\rm
eff}(\mu)>E_0$ for all $\mu$, then the sum rule (\ref{fit}) may be
solved for any $R$ within the broad range $0.7<R/R_0<1.15$
(Fig.~\ref{Fig:2}). Thus, the sum rule alone, without knowing the
continuum contribution, cannot determine $R$.

\begin{figure}[t]\begin{tabular}{cc}
\includegraphics[width=6.5cm]{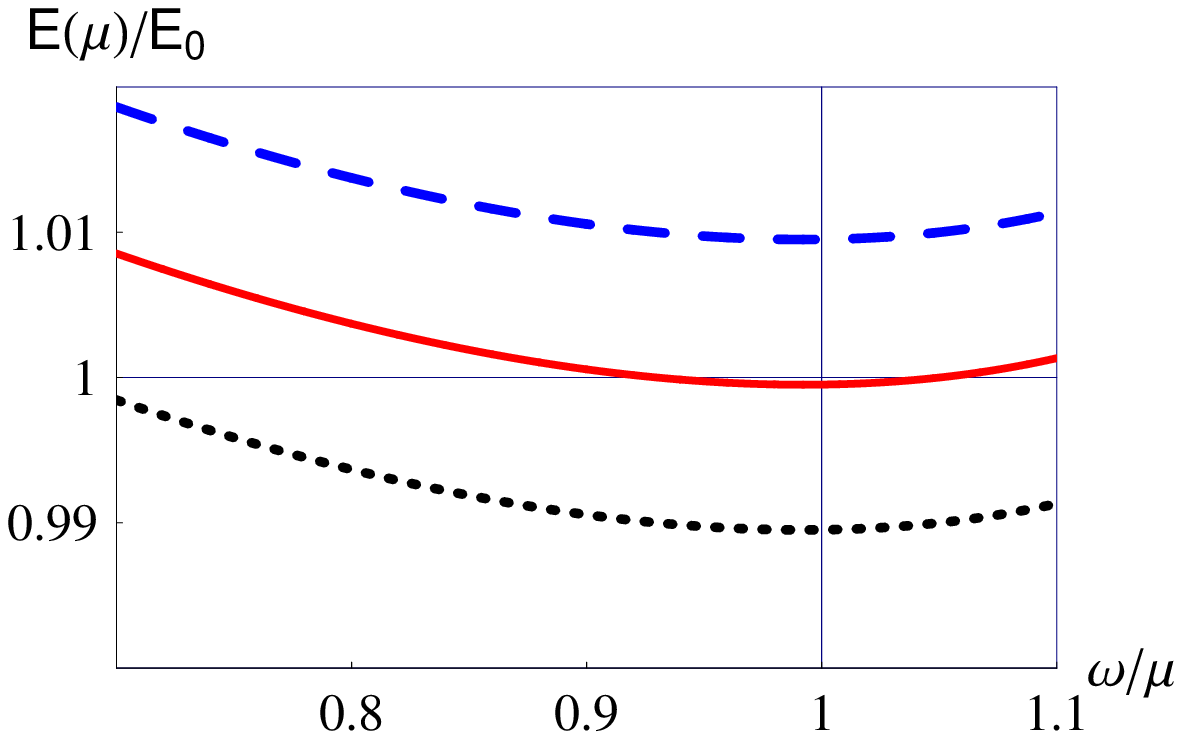}&\hspace{.2cm}
\includegraphics[width=6.5cm]{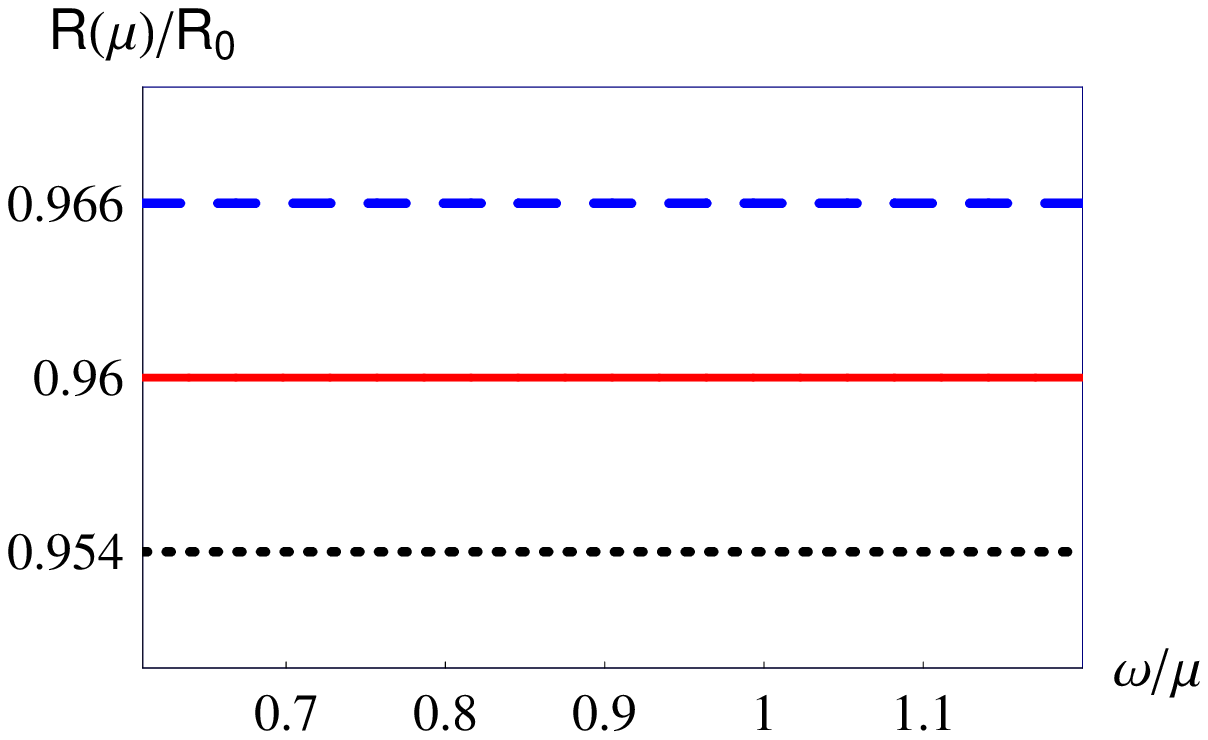}\\\end{tabular}
\caption{\label{Fig:3}Constant effective continuum threshold
$z_c$: $E(\mu)$ for three different values of $z_c$ (a) and the
corresponding $R(\mu)$ (b).}\end{figure}

\vspace{.2cm}\noindent\underline{\it Constant effective continuum
threshold.} Strictly speaking, a constant effective continuum
threshold $z_{\rm eff}(\mu)=z_c={\rm const}$ is incompatible with
the sum rule. Nevertheless, this Ansatz may work well, especially
in our model: as can be seen from Fig.~\ref{Fig:2}(a), the exact
$z_{\rm eff}(\mu)$ is almost flat in the fiducial interval.
Therefore, the HO model represents a very favorable situation for
applying the QCD sum-rule machinery.

Now, one needs to impose a criterion for fixing $z_c$. A widely
used procedure is the following \cite{jamin}: one calculates
\begin{eqnarray}
E(\mu,z_c)\equiv -\frac{d}{d(1/\mu)}\log \Pi(\mu,z_c),
\end{eqnarray}
which now depends on $\mu$ due to approximating $z_{\rm eff}(\mu)$
by some constant. Then, one determines $\mu_0$ and $z_c$ as the
solution to the system of equations
\begin{eqnarray}
\label{add} E(\mu_0,z_c)=E_0,\qquad
\left.\frac{\partial}{\partial\mu}E(\mu,z_c)\right|_{\mu=\mu_0}=0,
\end{eqnarray}
yielding $z_c=2.454\,\omega$, $\mu_0/\omega=1$ (Fig.~\ref{Fig:3}),
and a good central-value estimate $R/R_0=0.96$, with $R$ being
extremely stable in the fiducial range.

Note, however, a dangerous point: (i) a perfect description of
$\Pi(\mu)$ with better than 1\% accuracy, (ii) the deviation of
the $E(\mu,z_c)$ from $E_0$ at the level of only 1\%, (iii) and
extreme stability of $R(\mu)$ in the full fiducial range leads to
a 4\% error in the extracted value of $R$! Clearly, this error
could not be guessed on the basis of the other numbers obtained,
and it would be wrong to estimate the error, e.g., from the range
covered by $R$ when varying the Borel parameter $\mu$ within the
fiducial interval.

\vspace{.2cm}\noindent Let us summarize the lessons we have
learnt:

\noindent 1. The knowledge of the correlator in a limited range of
the Borel parameter $\mu$ is not sufficient for an extraction of
the ground-state parameters with a controlled accuracy: Rather
different models for the correlator in the form of a ground state
plus an effective continuum lead to the same correlator.

\noindent 2. The procedure of fixing the effective continuum
threshold by requiring the average mass calculated with the cut
correlator to reproduce the known ground-state mass
\cite{jamin,bz,lc_lms} is, not restrictive: a $\mu$-dependent
effective continuum threshold $z_{\rm eff}(\mu)$ which solves the
sum rule (\ref{fit}) leads to the cut correlator (\ref{cut}) which
automatically (i) reproduces exactly $E(\mu)=E_0$ for all values
of the Borel parameter $\mu$, and (ii) yields a $\mu$-independent
value of $R$ which, however, may be rather far from the true
value.

\noindent 3. Modeling the hadron continuum by a {\it constant}
effective continuum threshold allows one to determine $z_c$ by,
e.g., requiring the average energy $E(\mu)$ to be close to $E_0$
in the stability region. In the model under discussion (where the
exact effective continuum threshold is in fact almost constant)
one obtains in this way a good estimate $R/R_0=0.96$, with
practically $\mu$-independent $R$. The unpleasant feature is that
the deviation of $R$ from $R_0$ turns out to be much larger than
the variations of $E(\mu)$ and $R$ over the fiducial range.
Therefore, the systematic errors cannot be estimated by looking at
the variation of $R$ within the fiducial domain of $\mu$.

Thus, we conclude that in those cases where the continuum
contribution is not known (which is a typical situation
encountered, e.g., in analyses of heavy-meson observables) the
standard procedures adopted in a QCD sum-rule extraction of hadron
parameters do not allow to control the systematic uncertainties.
Consequently, no systematic errors for hadron parameters obtained
with sum rules can be provided. Let us emphasize that the
independence of the extracted values of hadron parameters from the
Borel mass does~not guarantee the extraction of their true values.

Nevertheless, in the model under consideration the sum rules give
good estimates for the parameter $R_0$. This may be due to the
following specific features of the model: (i) a large gap between
the ground state and the first excitation that contributes to the
sum rule; (ii) an almost constant exact effective continuum
threshold in a wide range of $\mu$. Whether or not the same good
accuracy may be achieved in QCD, where the features mentioned
above are absent, is not obvious, and requires more detailed
investigations.

This shortcoming --- the impossibility to control the systematic
errors --- remains the weak feature of the method of sum rules and
an obstacle for using the results from QCD sum rules for precision
physics, such as electroweak physics.

\vspace{.3cm}{\small{\it Acknowledgments.} We would like to thank
the organizers for the invitation to this interesting conference
and the Austrian Science Fund (FWF) for support under project
P17692.}

\end{document}